\documentclass[runningheads]{llncs}
\usepackage[T1]{fontenc}
\usepackage[hyphens]{url}
\usepackage{graphicx}
\usepackage{booktabs}
\usepackage{subcaption}
\usepackage{listings}
\usepackage[most]{tcolorbox}
\usepackage{todonotes}
\usepackage{caption}
\usepackage{varwidth}
\definecolor{myblue}{HTML}{118ab2}
\definecolor{myred}{HTML}{ef476f}
\usepackage{xcolor}
\usepackage{caption}
\definecolor{background}{HTML}{EEEEEE}
\definecolor{chatgptgreen}{HTML}{74AA9C}
\colorlet{numb}{magenta!60!black}

\definecolor{codebg}{rgb}{0.95, 0.95, 0.95}   
\definecolor{keyword}{rgb}{0.0, 0.0, 1.0}     
\definecolor{comment}{rgb}{0.0, 0.5, 0.0}     
\definecolor{string}{rgb}{0.58, 0.0, 0.82}    
\definecolor{number}{rgb}{0.0, 0.0, 0.5}      
\definecolor{punct}{rgb}{0.3, 0.3, 0.3}       
\definecolor{delim}{rgb}{0.2, 0.2, 0.2}       

\lstdefinestyle{customphp}{
  backgroundcolor=\color{codebg},
  basicstyle=\ttfamily\footnotesize,
  keywordstyle=\color{keyword}\bfseries,
  commentstyle=\color{comment}\itshape,
  stringstyle=\color{string},
  numberstyle=\color{number},
  breaklines=true,
  frame=single,
  keepspaces=true,
  numbers=left,
  numbersep=5pt,
  showspaces=false,
  showstringspaces=false,
  showtabs=false,
  tabsize=4,
  captionpos=b,
  language=PHP
}

\lstdefinelanguage{json}{
    basicstyle=\normalfont\ttfamily,
    numbers=none,
    showstringspaces=false,
    breaklines=true,
    frame=lines,
    backgroundcolor=\color{codebg},
    string=[s]{"}{"},
    stringstyle=\color{keyword},
    comment=[l]{:},
    commentstyle=\color{string},
}

\lstdefinestyle{customcsharp}{
  backgroundcolor=\color{codebg},
  basicstyle=\ttfamily\footnotesize,
  keywordstyle=\color{keyword}\bfseries,
  commentstyle=\color{comment}\itshape,
  stringstyle=\color{string},
  numberstyle=\color{number},
  breaklines=true,
  frame=single,
  keepspaces=true,
  numbers=left,
  numbersep=5pt,
  showspaces=false,
  showstringspaces=false,
  showtabs=false,
  tabsize=2,
  captionpos=b,
  language=[Sharp]C
}

\newcommand{\mybox}[2]{\fcolorbox{#1}{#1!10}{#2}}

\usepackage{float}
\newfloat{lstfloat}{htbp}{lop}
\floatname{lstfloat}{Listing}

\begin{document}
\title{The potential of LLM-generated reports in DevSecOps}

\author{Nikolaos Lykousas\inst{1}\orcidID{0000-0001-8874-1230} \and
Vasileios Argyropoulos\inst{1} \and Fran Casino\inst{1,2}\orcidID{0000-0003-4296-2876} 
}
%
\authorrunning{N. Lykousas et al.}

\institute{Data Centric, Romania \and
Department of Computer Engineering and Mathematics, Universitat Rovira i Virgili \\
\email{nlykousas@data-centric.eu}}
\maketitle              
\begin{abstract}
Alert fatigue is a common issue faced by software teams using the DevSecOps paradigm. The overwhelming number of warnings and alerts generated by security and code scanning tools, particularly in smaller teams where resources are limited, leads to desensitization and diminished responsiveness to security warnings, potentially exposing systems to vulnerabilities. This paper explores the potential of LLMs in generating actionable security reports that emphasize the financial impact and consequences of detected security issues, such as credential leaks, if they remain unaddressed. A survey conducted among developers indicates that LLM-generated reports significantly enhance the likelihood of immediate action on security issues by providing clear, comprehensive, and motivating insights. Integrating these reports into DevSecOps workflows can mitigate attention saturation and alert fatigue, ensuring that critical security warnings are addressed effectively.

\end{abstract}
\section{Introduction}
The adoption of DevSecOps pipelines through incorporating security tools for Static Application Security Testing (SAST)\cite{yang2019towards} and Dynamic Application Security Testing (DAST)\cite{lombardi2023devops} into the DevOps workflows, is a strategy aimed at identifying and mitigating bugs and vulnerabilities early in the development lifecycle, before the deployment. Nevertheless, many of the available security tools frequently produce inaccurate results or irrelevant results (e.g., false positives and negatives) \cite{johnson2013don}, requiring the software teams to manually assess such outputs. 
For example, poor quality of scanning results and limitations related to the outputs of security tools hinder rapid deployments, as manual and time-consuming tasks that require security expertise are needed to address these issues \cite{rajapakse2021empirical}.
Thus, despite the advantages of security scanning tools, developers often encounter significant challenges in effectively managing the influx of alerts and warnings generated by these workflows, especially if there are multiple tools involved, often providing overlapping results. This phenomenon, known as alert fatigue, is characterized by the overwhelming volume of notifications, leading to desensitization and diminished responsiveness to security warnings~\cite{ban2021combat}. This is especially problematic in smaller teams where developers have multiple responsibilities, making it challenging to prioritize and address every alert effectively. Studies show that false positive rates can exceed 50\%, contributing significantly to alert fatigue and the potential neglect of genuine vulnerabilities \cite{Sharif2023}.

At the same time, as management pushes for rapid deployment goals for software teams, developers often knowingly overlook identified security issues. This occurs because developers frequently fail to realize the full implications of a cyber attack or data breach on their organization, given that they often have limited liability and may not be directly affected by the consequences~\cite{Patrick2022}. Furthermore, the pressure to meet tight deadlines and deliver features quickly can lead developers to prioritize functionality over security, exacerbating the problem. Despite the rapid growth of AI-based automated solutions for code vulnerability assessment \cite{Moumita2023software,guo2024esorics}, critical vulnerabilities in code that end up in production remain unaddressed, posing significant risks to the affected organizations. These risks include financial losses, reputational damage, regulatory penalties, and operational disruptions. For instance, high-profile breaches such as the Target data breach in 2013~\cite{Dyess2016} and the Equifax breach in 2017~\cite{zou2018ve} demonstrate the severe consequences of unaddressed security vulnerabilities. These incidents resulted in millions of dollars in losses, legal actions, and long-lasting reputation damage.

This paper explores the potential of LLMs in generating detailed, actionable security reports from code scanning tool outputs, which are often either machine-readable (e.g. JSON) and not easily interpretable by humans, or overly verbose and difficult to digest. Beyond highlighting the issues provided by the tools, LLM-generated reports can assess the financial impact and potential consequences of not addressing the specific vulnerabilities if they were to be exploited by an adversary. We hypothesize that this approach can strongly motivate developers to act on patching vulnerabilities instead of overlooking them.

To this end, we conducted a survey ($N=23$) to evaluate the effectiveness of LLM-generated reports in enhancing security awareness and prompt action. The survey assesses several dimensions, including the clarity of the reports, the comprehensiveness of the information regarding potential consequences provided, the perceived relevance of financial impact assessments, and the overall motivational impact on respondents.

\section{Background}
\subsection{DesvSecOps}
DevSecOps is a recent paradigm shift in software engineering, aimed at integrating security practices into the DevOps framework. DevOps, initially focused on improving collaboration between development and operations teams, has evolved to include security, leading to the emergence of DevSecOps. This transformation is driven by the need to address security concerns that have become increasingly significant with the rise of continuous integration and continuous delivery (CI/CD) practices \cite{sanchez2020security,myrbakken2017devsecops}.

Essentially, DevSecOps represents embedding security controls and processes into every phase of the software development lifecycle, promoting collaboration among development, operations, and security teams in medium to large size organizations~\cite{sanchez2020security}, or proactively notifying developers of possible vulnerabilities and issues before pushing code into production for smaller teams. The integration of security into DevOps, means including security considerations early in the development process, rather than addressing them only at the final stages~\cite{myrbakken2017devsecops}. 

\subsection{LLM Report Generation}
LLMs have the ability to process large amounts of data and generate human-like text, and they have proven their value in domains such as healthcare and legal services. LLMs can offer diagnostic suggestions, legal advice, and even write comprehensive reports summarizing large contexts, bridging gaps where human expertise may be limited or unavailable. In the domain of medical diagnosis, LLMs have shown promise in analyzing multimodal data—combining medical images and symptom descriptions to provide diagnostic reports \cite{panagoulias2023evaluating}. Moreover, in the context of advice-seeking, studies indicate that LLM-generated advice can influence user behavior~\cite{wester2024exploring}.

\section{Methodology}
In this section, we present our approach to report generation with LLMs. For input, we opt for the LAZARUS AI Secret Scanner tool,  a novel language-agnostic secret scanning solution presented in \cite{lykousas2023tales,lykousas2024decoding}, capable of identifying hardcoded credentials in source code, either human-generated (passwords), or machine-generated (API keys/secrets, tokens, etc.). Different than popular SAST tools with similar functionality, such as TruffleHog\footnote{\url{https://github.com/trufflesecurity/truffleHog}} which follow the conventional approach to secret detection, i.e. looking for high entropy strings or using regular expressions, LAZARUS AI Secret Scanner employs an ensemble of two text convolution neural network models based on TextCNN architecture \cite{zhang2015character}, namely the Context Model and the Password Model. The Context Model is trained to classify source code snippets surrounding specific seed elements (including names of methods, variables, constants, etc.) that can be potentially relevant for various authentication contexts, including databases, mail servers, automation, web services and more. The Password Model is trained to classify strings extracted for the code snippets previously identified as potential passwords, machine-produced secrets (including API keys, JWT tokens, etc.), or ordinary strings. From experiments conducted in a large-scale dataset, LAZARUS AI Secret Scanner has shown compelling performance in terms of both correctly identifying credentials and the type of authentication context they appear in.

\begin{figure}[t!]
\begin{lstlisting}[style=customphp]
<?php
define('API_KEY', 'sk-*****8wLeuT3BlbkFJy9mYE9ogfUII9wM4Vvih');
$conn = mysqli_connect("mysql-***.alwaysdata.net", "276456_admin", 
"B34ut1fu1223@", "gitproject");
if($conn === false){
    die("ERROR: Could not connect: ". mysqli_connect_error());
}
?>
\end{lstlisting}

\begin{lstlisting}[language=json]
{
    "file": "***.php",
    "secrets": [
        {
            "match": "sk-*****8wLeuT3BlbkFJy9mYE9ogfUII9wM4Vvih",
            "line": 2,
            "classification": "OpenAI Secret"
        }
    ],
    "passwords": [
        {
            "match": "B34ut1fu1223@",
            "type": "argument",
            "line": 3,
            "username": "276456_admin",
            "classification": "Database"
        }
    ]
}
\end{lstlisting}
\caption{Example source code file with hardcoded credentials, and the corresponding output of LAZARUS AI Secret Scanner.} 
\label{fig:code_output}
\end{figure}

Moreover, we consider the large-scale dataset, presented in \cite{lykousas2023tales}, which comprises more than 2 million secrets that were committed in public repositories on GitHub. This dataset contains labels according to programming language, type of secret (e.g., token, password), usage (e.g., mail, database), etc. From this dataset, we randomly select 10 files containing hardcoded credentials in both categories supported by the tool (human generated/machine generated), and use LAZARUS AI Secret Scanner. A sample of the relevant authentication context and the tool's JSON-formatted output is presented in Figure~\ref{fig:code_output}.

\begin{table}[th]
    \centering
    \caption{Instructions for the report-generation task.}
    \begin{subtable}[h]{\textwidth}
        \centering
        \begin{tabular}{|p{.95\textwidth}|}
            \hline \textbf{Prompt:} Given the hardcoded secrets and credentials detected by a SAST source code scanner tool, generate a concise report analysing the potential consequences and the estimated financial impact to an organisation, if the affected source code is released into production, or accessed by an adversary. The generated report is aimed to developers and technical managers, and needs to motivate them to take action. Do not repeat the provided source code. \\
            Secret Scanner output: \texttt{\textasciigrave\textasciigrave\textasciigrave}\mybox{myblue}{[JSON]}\texttt{\textasciigrave\textasciigrave\textasciigrave} \\
            Affected source code: \texttt{\textasciigrave\textasciigrave\textasciigrave}\mybox{myblue}{[CODE]}\texttt{\textasciigrave\textasciigrave\textasciigrave} \\
            \hline
        \end{tabular}
    \end{subtable}~
    
    \label{tbl:tasks}
\end{table}

Next, we consider two state-of-the-art LLMs at the time of writing, one proprietary and one open-source one, namely OpenAI's ChatGPT\footnote{\texttt{gtp-4o-2024-05-13}} and Meta's Llama 3\footnote{\texttt{llama-3-70b-instruct}}. For the selected files, we combine their source code and the tool outputs, and prompt the LLMs using the template defined in Table~\ref{tbl:tasks}. We present a sample of outputs from each model in Figure \ref{fig:llm_reports}.

\begin{tcolorbox}[colback=chatgptgreen!7, colframe=chatgptgreen, boxrule=0.5mm, width=\textwidth, arc=2mm, auto outer arc, enlarge left by=0mm, enlarge right by=0mm, enlarge top by=2mm, enlarge bottom by=2mm, title=ChatGPT, breakable, fonttitle=\bfseries, coltitle=black, colbacktitle=white, fontupper=\small, floatplacement=H]
\textbf{Security Incident Report: Hardcoded Secrets Detected in Source Code}

A static application security testing (SAST) tool has detected hardcoded secrets in a PHP file. This report outlines the potential consequences and financial impact if the affected source code is released into production or accessed by an adversary. The goal is to motivate developers and technical managers to take immediate action to remediate the identified vulnerabilities.

\textbf{Potential Consequences}
\textbf{1. Compromised OpenAI API Key}
\begin{itemize}
    \item \textbf{Service Disruption:} Unauthorized use of the OpenAI API could lead to service disruptions, including rate limiting or service suspension due to abuse.
    \item \textbf{Financial Loss:} Excessive usage of the OpenAI API by malicious actors could incur significant costs.
    \item \textbf{Data Exposure:} Sensitive data processed through the OpenAI API could be intercepted or exploited by unauthorized users.
\end{itemize}

\textbf{2. Compromised Database Credentials}
\begin{itemize}
    \item \textbf{Data Breach:} Unauthorized access to the database could lead to a breach of sensitive information, including user data, intellectual property, and operational data.
    \item \textbf{Service Disruption:} Malicious actors could alter or delete database records, leading to operational disruptions.
    \item \textbf{Financial Impact:} The costs associated with data breaches include notification expenses, regulatory fines, and loss of customer trust.
\end{itemize}

\textbf{Estimated Financial Impact}

\textbf{Data Breach Cost Estimation}
\begin{itemize}
    \item \textbf{Average Cost per Record:} According to IBM's 2023 Cost of a Data Breach Report, the average cost per compromised record is approximately \$164.
    \item \textbf{Estimated Number of Records:} Assuming the database contains around 10,000 records, the total estimated cost would be \$1,640,000.
\end{itemize}

\textbf{API Abuse Cost Estimation}
\begin{itemize}
    \item \textbf{Potential Monthly Overuse:} If a malicious actor makes 100,000 requests to the OpenAI API, assuming a cost of \$0.02 per request, the monthly cost could reach \$2,000.
    \item \textbf{Annual Cost Estimate:} If unchecked for a year, the potential cost would be \$24,000.
\end{itemize}

\textbf{Total Financial Impact}
Combining the potential costs of a data breach and API abuse, the total estimated financial impact could exceed \$1,664,000.

\textbf{Recommendations}

\textbf{Immediate Actions}
\begin{itemize}
    \item \textbf{Revoke and Rotate Secrets:} Immediately revoke the exposed OpenAI API key and database credentials. Generate and use new secrets.
    \item \textbf{Environment Variable Usage:} Refactor the code to use environment variables for storing sensitive information instead of hardcoding them.
    \item \textbf{Audit and Monitor:} Conduct a thorough audit of the codebase to identify any other hardcoded secrets. Implement monitoring tools to detect unauthorized access or usage.
\end{itemize}

\textbf{Long-Term Measures}
\begin{itemize}
    \item \textbf{Implement Secret Management Solutions:} Use secret management tools (e.g., HashiCorp Vault, AWS Secrets Manager) to securely manage and access sensitive information.
    \item \textbf{Continuous Security Scanning:} Integrate SAST tools into the CI/CD pipeline to detect and remediate vulnerabilities early in the development process.
    \item \textbf{Developer Training:} Provide training for developers on secure coding practices and the importance of protecting sensitive information.
\end{itemize}

\textbf{Conclusion}
The detected hardcoded secrets in the source code pose significant security risks and potential financial impact. Immediate action is required to mitigate these risks and prevent potential breaches or abuses. Implementing the recommended measures will enhance the security posture and safeguard the organization from future incidents.

\textbf{Call to Action}
Developers and technical managers must prioritize the remediation of these vulnerabilities to protect the organization’s assets and maintain customer trust. The cost of inaction far outweighs the effort required to secure the codebase. For further assistance or to discuss the next steps, please contact the security team.
\end{tcolorbox}
\begin{tcolorbox}[colback=blue!7, colframe=blue!40!black, boxrule=0.5mm, width=\textwidth, arc=2mm, auto outer arc, enlarge left by=0mm, enlarge right by=0mm, enlarge top by=2mm, enlarge bottom by=2mm, title=Llama 3, breakable, fonttitle=\bfseries, coltitle=black, colbacktitle=white, fontupper=\small, floatplacement=H]
\textbf{Urgent: Hardcoded Secrets and Credentials Detected in Source Code}

\textbf{Summary:}
A SAST source code scanner tool has identified hardcoded secrets and credentials in the \texttt{***.php} file, which poses a significant security risk to our organization. If this code is released into production or accessed by an adversary, it could lead to unauthorized access to sensitive data, systems, and financial losses.

\textbf{Potential Consequences:}
\begin{enumerate}
    \item \textbf{Unauthorized Access to OpenAI Services:} The exposed OpenAI secret key (\texttt{sk-*****8wLeuT3BlbkFJy9mYE9ogfUII9wM4Vvih}) could be used by an attacker to access our OpenAI account, leading to unauthorized usage, data breaches, or malicious activities.
    \item \textbf{Database Compromise:} The hardcoded database credentials (\texttt{B34ut1fu1223@} and \texttt{276456\_admin}) could allow an attacker to gain unauthorized access to our database, resulting in data theft, tampering, or destruction.
    \item \textbf{Lateral Movement:} An attacker could use the compromised database access to move laterally within our network, potentially gaining access to other systems, data, and sensitive information.
    \item \textbf{Run Malicious Code:} They could leverage our quota to generate harmful content or launch denial-of-service attacks against other users.
\end{enumerate}

\textbf{Estimated Financial Impact:}
\begin{itemize}
    \item \textbf{Data Breach Costs:} The average cost of a data breach is estimated to be around \$3.92 million (according to IBM's 2020 Cost of a Data Breach Report). In this scenario, the exposure of sensitive data could lead to significant financial losses.
    \item \textbf{System Downtime and Recovery:} In the event of a successful attack, the downtime and recovery costs could be substantial, potentially exceeding \$100,000 per day.
    \item \textbf{Reputation Damage:} A security breach could damage our organization's reputation, leading to a loss of customer trust, revenue, and long-term consequences.
\end{itemize}

\textbf{Recommendations:}
\begin{enumerate}
    \item \textbf{Immediately Remove Hardcoded Secrets and Credentials:} Update the code to use secure storage mechanisms, such as environment variables, secure storage services, or encrypted files.
    \item \textbf{Implement Secure Coding Practices:} Ensure that all developers follow secure coding guidelines to prevent similar vulnerabilities in the future.
    \item \textbf{Conduct a Comprehensive Security Audit:} Perform a thorough security audit to identify and address any other potential vulnerabilities in our codebase.
\end{enumerate}

\textbf{Action Required:}
We urge all developers to take immediate action to address this critical security issue. Failure to do so could result in significant financial losses, reputational damage, and legal consequences.

Please prioritize the removal of hardcoded secrets and credentials, and implement secure coding practices to prevent similar vulnerabilities in the future. If you have any questions or concerns, please do not hesitate to reach out to the security team.

\end{tcolorbox}
\vspace{-2em} 
\captionsetup{type=figure}
\captionof{figure}{LLM-generated reports}
\label{fig:llm_reports}
\vspace{1em}

From the LLM outputs, it can be observed that in every case, they returned sensible comments regarding the consequences of a possible compromise involving the detected credential and the provided authentication context. Interestingly, both models also provided reasonable mitigation recommendations, even without being explicitly prompted. However, ChatGPT consistently provided more detailed and actionable recommendations. Additionally, we observed that some of the possible consequences and recommendations generated by Llama were not realistic. One such example can be seen in Figure~\ref{fig:llm_reports} (i.e., refer to the \texttt{Run Malicious Code} as described in the \textit{Potential Consequences} summary). Moreover, as expected the financial impact figures reported are hallucinations, and not always reasonable. In particular, ChatGPT provided significantly high costs, which would only be relevant for breaches in every large enterprises, and not always aligned with the context of the provided code. On the other hand, Llama often omitted precise cost estimates. Notably, in most of the reports, both models cited the IBM Cost of Data Breach Report\footnote{\url{https://www.ibm.com/reports/data-breach}}, though the reported numbers were hallucinated. Nevertheless, in this case, there is not a significant deviation from the real data. Finally, we observed that, in general, the reports generated by ChatGPT were more impersonal, while Llama 3 often used first-person language (e.g. the pronoun "our") in the generated reports. This choice of wording can simulate a human-like and motivational tone, making the communication feel more engaging~\cite{tsai2021chatbots}.

\section{Survey Design and Results}
After collecting the LLM-generated reports for each file in our sample, we created a survey to  evaluate how effectively these reports motivate developers to take action on resolving identified security issues, and how this compares to the reports or alerts generated by the security tools in their DevSecOps pipelines. To this end, the survey focuses on the clarity of consequences/perceived financial impact, and overall motivation for action compared to the status quo of alerting and reporting in the participants' workflows. 

As such, we sent the generated reports (20 in total, 1 for reach LLM) and the source code samples to 23 software engineers who agreed to participate in the survey. The participants were selected on the basis that they already employ DevSecOps workflows and have experience with SAST and similar security tools for source code auditing. Moreover, we created a short questionnaire of 5 questions that participants were asked to respond to for each report, with answers aligned with Likert-scale~\cite{likert1932technique}. The questions are presented in Table~\ref{table:survey_questions}.

\begin{table}[ht!]
\centering
\caption{Survey questions \& possible answers for evaluating the LLM-generated reports}
\begin{tabular}{|p{\textwidth}|}
\hline
\textbf{Survey Questions} \\ \hline
\textbf{Q1}. How clear and reasonable are the consequences of the identified security issue(s) in the report? \\
\textbf{Answers:} Very unclear / Unclear / Neutral / Clear / Very clear \\ \hline
\textbf{Q2}. How effectively does the report communicate the financial impact of the security issue(s)? \\
\textbf{Answers:} Very ineffective / Ineffective / Neutral / Effective / Very effective \\ \hline
\textbf{Q3}. How motivated are you to take action based on the report? \\
\textbf{Answers:} Not motivated at all / Slightly motivated / Neutral / Motivated / Very motivated \\ \hline
\textbf{Q4}. How does the motivational impact of the report compare to the alerts and reports generated by your current DevSecOps tools? \\
\textbf{Answers:} Much worse / Worse / About the same / Better / Much better \\ \hline
\textbf{Q5}. How much do you trust the recommendations provided in the report? \\
\textbf{Answers:} Not at all / Slightly / Neutral / Mostly / Completely \\ \hline
\end{tabular}

\label{table:survey_questions}
\end{table}

\begin{figure*}[th!]
   \begin{subfigure}[t]{\columnwidth}
	\centering
      \includegraphics[width=\columnwidth]{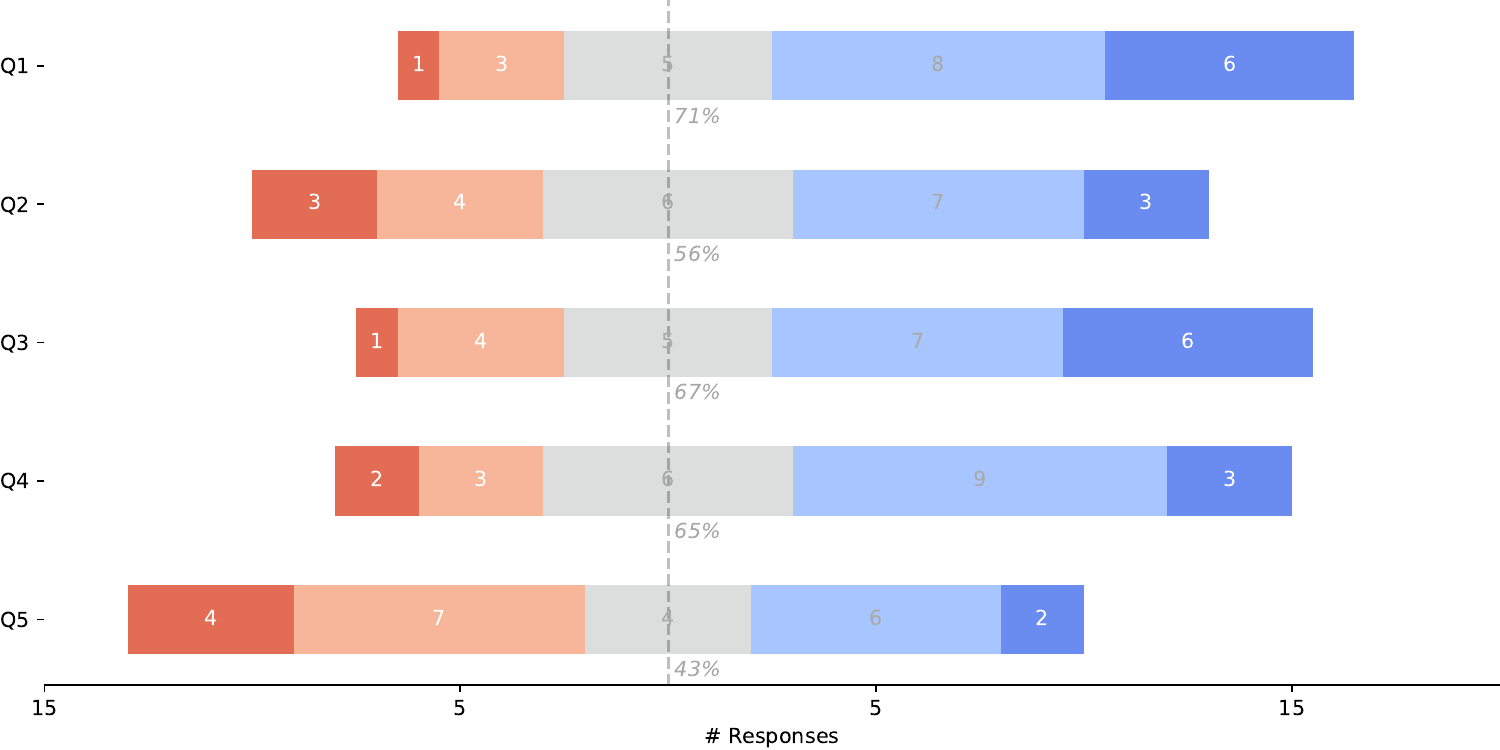}
      \caption{ChatGPT reports}
      
   \end{subfigure}
   \hfill
   \begin{subfigure}[t]{\columnwidth}
	\centering
	\includegraphics[width=\columnwidth]{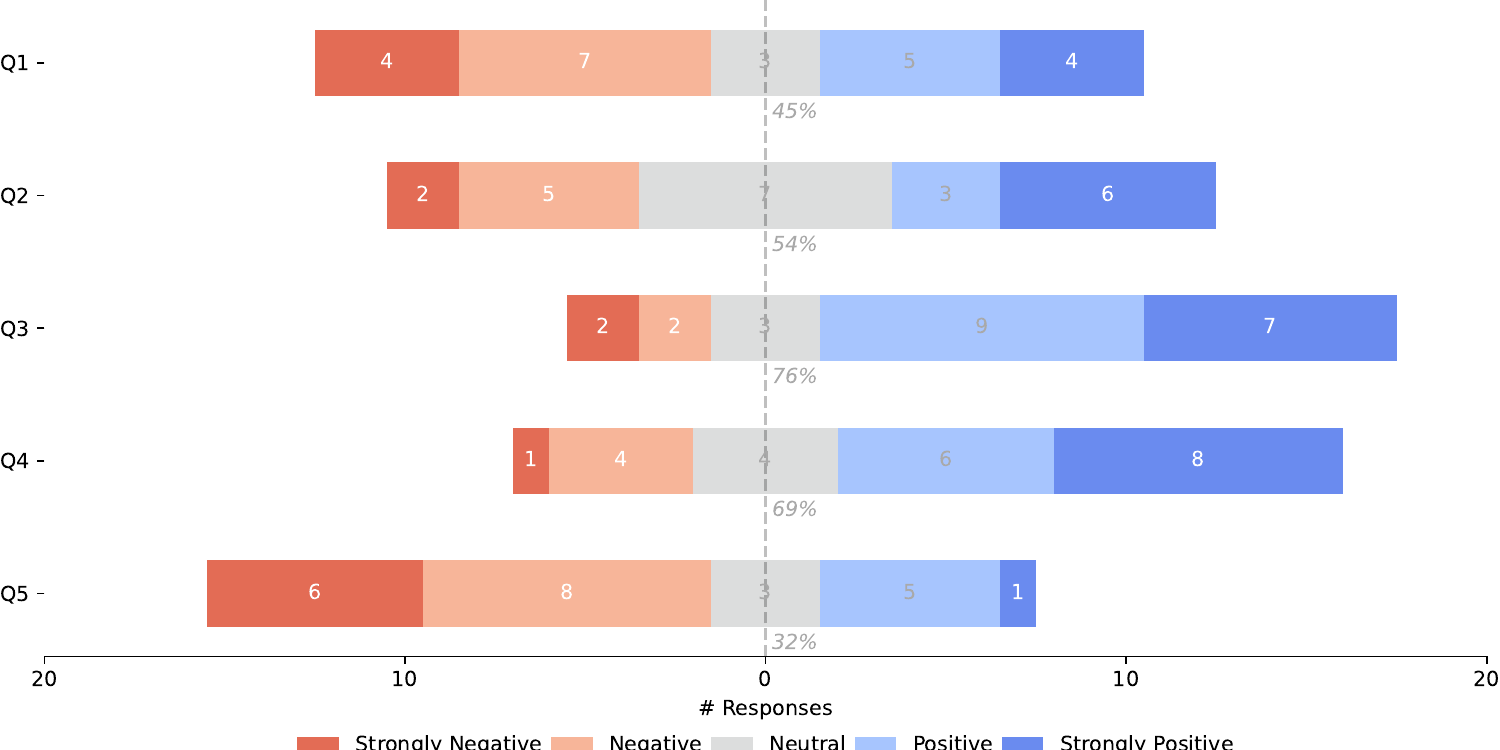}
    \caption{Llama reports}	
	\end{subfigure}
  \caption{Answers to the survey per question and LLM in Likert-scale.}
   \label{fig:answers}
\end{figure*}

Next, we present the results to the questionnaire in Figure~\ref{fig:answers} for each LLM. For ChatGPT, the majority of participants found the consequences of the identified security issues to be clear and reasonable, with 8 positive and 6 very positive responses. Additionally, ChatGPT was perceived as effective in communicating the financial impact, though there were some neutral and slightly negative responses, suggesting room for improvement in conveying this aspect more convincingly, provided that as previously noted, the reported numbers tend to be exaggerated. Llama 3, on the other hand, showed a mixed response regarding the clarity of consequences, with a significant portion of participants disagreeing or remaining neutral. 

Nevertheless, the motivation to take action based on LLM-generated reports is high for both models. For ChatGPT, 7 participants were motivated, and 6 were very motivated. For Llama 3, 9 were motivated, and 7 were very motivated. This indicates that the reports have the potential to motivate the developers to take action in terms of addressing the identified issues.
Notably, the more personal style of Llama reports  appears to be more effective in persuading developers to take action.

Conversely, trust in the recommendations showed a mix of moderate agreement and significant skepticism for both models. Although the recommendations in all cases were reasonable and aligned with the best practices in DevSecOps, developers seem biased against the trustworthiness of LLMs in critical contexts like cybersecurity. This reflects a broader challenge associated with trust of LLM-generated content.

\section{Conclusions}
The integration of LLM-generated reports into DevSecOps workflows shows significant potential in addressing the challenges of alert fatigue and motivating developers to take prompt action on security issues. The survey results indicate that both ChatGPT and Llama effectively communicate the potential consequences and financial impacts of identified security vulnerabilities, with a majority of participants finding the reports clear and actionable. Notably, ChatGPT excelled in providing clear and reasonable explanations of security issues, while Llama 3's more personal and motivational tone proved effective in driving action among developers. In both cases, developers found the LLM-generated reports more motivating to act on the identified issues compared to the alerts and reporting of security issues in their current workflows. 

Despite their advantages, the idea of LLMs generating reports in the DevSecOps context faces challenges in gaining trust for their recommendations. This could be attributed to the generic nature of the described solutions, which, although aligned with security best practices, might not always be relevant in tackling specific issues, such as credential leaks. Additionally, the developers' responses indicate a broader skepticism towards LLMs in critical security contexts, reflecting doubts about their reliability and accuracy. Nonetheless, the survey responses indicating a preference for Llama in producing motivating reports suggest that there is significant potential for integrating local, open-source LLMs into developers' workflows without the need for their code to leave their premises. Beyond coding assistance, such models could be used for alerting and report generation, mitigating alert fatigue with their capacity to produce compelling reports, and thus ensuring that critical security warnings do not fly under the radar of developers and security teams.

\begin{credits}
\subsubsection{\ackname} This work was supported by the European Commission under the Horizon Europe Programme, as part of the project LAZARUS (\url{https://lazarus-he.eu/}) (Grant Agreement no. 101070303). This research is supported by supported by Ministerio de Ciencia, Innovación y Universidades, Gobierno de España (Agencia Estatal de Investigación, Fondo Europeo de Desarrollo Regional -FEDER-, European Union) under the research grant PID2021-127409OB-C33 CONDOR. Fran Casino was supported by the Government of Catalonia with the Beatriu de Pinós programme (Grant No. 2020 BP 00035), and by AGAUR with the project ASCLEPIUS (2021SGR-00111).

The content of this article does not reflect the official opinion of the European Union. Responsibility for the information and views expressed therein lies entirely with the authors.

\subsubsection{\discintname}
The authors have no competing interests to declare that are relevant to the content of this article.
\end{credits}

\bibliographystyle{splncs04}
\bibliography{refs}
\end{document}